\documentclass[aps,prl,showpacs,nobibnotes,twocolumn,
nobalancelastpage,superscriptaddress]{revtex4}
\usepackage{graphicx}
\usepackage{amsmath}

\begin{document}


\title{Gamma-Ray Constraint on Galactic Positron Production by MeV Dark Matter}

\author{John F. Beacom}
\affiliation{Department of Physics, The Ohio State University,
Columbus, Ohio 43210} 
\affiliation{Department of Astronomy, The Ohio State University,
Columbus, Ohio 43210}

\author{Nicole F. Bell}
\affiliation{NASA/Fermilab Astrophysics Center, Fermi National
Accelerator Laboratory, Batavia, Illinois 60510-0500}
\affiliation{Kellogg Radiation Laboratory,
California Institute of Technology, Pasadena, California 91125}

\author{Gianfranco Bertone}
\affiliation{NASA/Fermilab Astrophysics Center, Fermi National
Accelerator Laboratory, Batavia, Illinois 60510-0500}

\date{16 September 2004}

\begin{abstract}
The Galactic positrons, as observed by their annihilation gamma-ray
line at 0.511~MeV, are difficult to account for with astrophysical sources.
It has been proposed that they are produced instead by dark matter
annihilation or decay in the inner Galactic halo.  To avoid other
constraints, these processes are required to occur ``invisibly," such
that the eventual positron annihilation is the only detectable signal.
However, electromagnetic radiative corrections to these processes
inevitably produce real gamma rays (``internal bremsstrahlung"); this
emission violates COMPTEL and EGRET constraints unless the dark matter
mass is less than about 20 MeV.
\end{abstract}


\pacs{95.35.+d, 98.70.Rz, 98.70.Sa
\hspace{2.1cm} FERMILAB-PUB-04-209-A, KRL-MAP-301}

\maketitle


{\bf Introduction.---} 
The SPI camera on the INTEGRAL satellite has recently observed the
0.511 MeV gamma-ray emission line arising from positron annihilation in
the Galaxy~\cite{Jean03,Knodlseder03,Weidenspointner04}.  The flux
from the Galactic center region is $\Phi_{511} = (9.9^{+4.7}_{-2.1})
\times 10^{-4}$ photons cm$^{-2}$ s$^{-1}$, confirming earlier
measurements~\cite{Johnson73,Leventhal78,Purcell97}.  The INTEGRAL
data also provide new clues on the morphology of the emission region,
which is consistent with a 2-dimensional gaussian of full width at
half maximum (FWHM) of 9 degrees, with a $2-\sigma$ uncertainty range
covering $6-18$ degrees.  

A variety of astrophysical sources of positrons have been proposed,
among them compact objects, massive stars, supernovae, gamma-ray
bursts, and cosmic rays (see, e.g.,
Refs.~\cite{Chan93,Milne02,Dermer01,Casse03,Bertone04b}).  However,
these astrophysical sources have difficulty accounting for the
intensity of the positron annihilation flux, and especially for the
morphology of the emission region, which so far in the INTEGRAL data
shows neither a disk component nor discrete sources.  These facts
motivate consideration of an exotic mechanism for the positron
production, and one associated with the dark matter concentration at
the Galactic center is naturally suggested.

However, for dark matter candidates in the usually considered range of
masses, 10 GeV--10 TeV, the production of positrons by dark matter
annihilation would be accompanied by the production of other
kinematically allowed particles (for a recent review of dark matter
candidates, see Ref.~\cite{Bertone04a}).  Even if direct production of
gamma rays were suppressed, there would be a gamma-ray flux arising
from the decays of secondaries (e.g., $\pi^0$, which are created in
the hadronization of quarks).  For annihilation of typical dark
matter candidates to account for the observed positrons, the
associated flux of high-energy gamma rays in the direction of the
Galactic center would exceed the EGRET data by several orders of
magnitude.

Boehm {\it et al.}~\cite{Boehm04a} have recently proposed that all of
the characteristics of the observed signal could be well fit by a
scenario in which {\it light} ($1-100$ MeV) dark matter particles
annihilate only into $e^+ e^-$ pairs; the rate is controlled by the
dark matter annihilation cross section (and its velocity dependence),
and the morphology by the assumed dark matter density
profile~\cite{Boehm04a,Boehm04b,Hooper03}.  Since the mass is below
100 MeV, the only other kinematically allowed annihilations would be
to gamma rays and neutrinos.  To avoid the direct gamma-ray
constraints (among them the cosmic gamma-ray
background~\cite{Zhang04}), and to account for the required positron
production rates, these modes are postulated to not occur.  While the
assumed dark matter mass is quite low, it is claimed that this model
is consistent with all laboratory
data~\cite{Boehm02,Boehm03,Hooper03,Boehm04a,Boehm04b}.  In addition,
the model is apparently consistent with astrophysical data and
big-bang nucleosynthesis, depending on the assumptions made about the
masses and couplings~\cite{Serpico04}.

In the model of Boehm {\it et al.}, the positrons have a seemingly
``invisible" birth, in that the dark matter annihilation only produces
$e^+ e^-$ pairs.  Another clever aspect of this model is that
positrons below 100 MeV will lose energy dominantly by ionization
(avoiding production of possibly detectable inverse Compton and
synchrotron radiation~\cite{Strong98a}) and will remain confined to
the Galactic center region (required to reproduce the observed
emission region).  The positrons thus remain invisible until their
annihilation into 0.511 MeV gamma rays.  Very similar considerations
apply to models of dark matter decay which produce the observed
positrons~\cite{Picciotto04,Hooper04}.

However, the dark matter annihilation process $\chi \chi \rightarrow
e^+ e^-$ is necessarily accompanied by the process $\chi \chi
\rightarrow e^+ e^- \gamma$, arising from electromagnetic radiative
corrections.  This real gamma-ray emission, illustrated in Fig.~1, is
known as {\it internal bremsstrahlung}, the name indicating that it
arises from the Feynman diagram itself and is not due to propagation
in a medium.  In addition, the flux and spectrum of the internal
bremsstrahlung gamma rays can be calculated with adequate accuracy
without knowing the new particle physics which mediates the dark
matter annihilation.  {\it The internal bremsstrahlung gamma rays
reveal the otherwise invisible dark matter annihilations, directly
testing the central assumption of the proposed models, which is that
the observed positrons were originally produced with energies up to
100 MeV}.  Astrophysical models would have to produce the same number
of positrons, but at lower energies, typically a few MeV.

We will show that for most of the proposed dark matter mass range, the
flux of internal bremsstrahlung gamma rays would be inconsistent with
COMPTEL and EGRET measurements of {\it diffuse} radiation from the
Galactic center region, thus requiring the dark matter mass to be less
than about 20 MeV.  This result is almost completely independent of
assumptions about the physical conditions at the Galactic center or
the new particle physics which mediates the dark matter annihilation.
In fact, it is more general than just dark matter annihilation or
decay, and prohibits any model for the positron production in which
the positrons are created at energies above about 20 MeV.  We will
also show that future Galactic gamma-ray data will be able to
significantly improve the sensitivity of our constraint.

\begin{figure}[t]
\includegraphics[width=3.25in,clip=true]{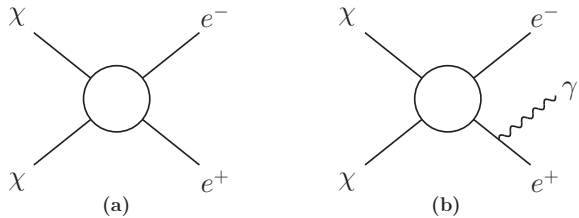}
\caption{\label{bremdiagram} Diagram (a) shows the dark matter
annihilation channel $\chi \chi \rightarrow e^+ e^-$ assumed by Boehm
{\it et al.}~\cite{Boehm04a}; the positron production and propagation
occur invisibly, with only the eventual positron annihilation being
detectable.  The open circle represents unspecified new physics.
Diagram (b) shows the observation of this paper, that positron
production must be accompanied by detectable internal bremsstrahlung
(a similar diagram with radiation from the electron is not shown).}
\end{figure}


{\bf Internal Bremsstrahlung.---}
The tree-level annihilation process $\chi \chi \rightarrow e^+ e^-$
will be subject to radiative corrections, which will affect the total
cross section at ${\cal O}(\alpha)$, where $\alpha = 1/137$ is the
fine-structure constant.  These corrections are expected to be
model-dependent but small.  The portion of the radiative corrections
that governs the emission of real gamma rays, i.e., the finite part of
the outer radiative corrections, is also expected to be small, but is
process-independent and factorizes from tree-level cross
section~\cite{Peskin95}.  Generally, those results apply to soft
gamma-ray radiation, but we will show that they are adequately
accurate for our purposes, even for large gamma-ray energies.
Internal bremsstrahlung has been observed in a very wide variety of
processes (one with similar positron energies is muon
decay~\cite{Crittenden61}).

The process $\chi \chi \rightarrow e^+ e^-$ from $\chi$ nearly at rest
(typical halo velocities are $10^{-3} c$) produces monoenergetic
electrons and positrons, with energies equal to the dark matter mass
$m_\chi$.  But when the gamma-ray energy is comparably large, it
shares the phase space available to the final state.  In order to
reduce the process dependence, we use the final-state internal
bremsstrahlung probability from the kinematically similar process $e^+
e^- \rightarrow \mu^+ \mu^-$~\cite{Martyn90,Berends98}, modified by
replacing the muon mass with the electron mass (since the initial
state is $\chi \chi$, it is appropriate to make this substitution
while ignoring the identical-particle issue in $e^+ e^- \rightarrow
e^+ e^-$).

The internal bremsstrahlung cross section is then
\begin{equation}
\frac{d\sigma_{\rm Br}}{dE} = \sigma_{\rm tot} \times \frac{\alpha}{\pi}
\; \frac{1}{E}
\left[ \ln \left( \frac{s'}{m_e^2} \right) -1 \right]   
\left[ 1 + \left( \frac{s'}{s} \right)^2 \right]\,,
\label{cross}
\end{equation}
where $E$ is the gamma-ray energy, and $\sigma_{\rm tot}$ is the
tree-level cross section, which factors out, an essential point.  Here
$s = 4 m_\chi^2$ and $s' = 4m_\chi (m_\chi - E)$.  When $m_\chi \gg E
, m_e$,
\begin{equation}
\frac{d\sigma_{\rm Br}}{dE} \simeq \sigma_{\rm tot} \times
\frac{4 \alpha}{\pi} \; \frac{\ln \left(2 m_\chi / m_e\right)}{E}\,,
\label{crossapprox}
\end{equation}
which displays the familiar scaling factors.  The internal
bremsstrahlung probability is very small for the gamma-ray energy
range which we consider, which is well above the soft singularity. We
find the most stringent constraints when E is large, near $m_\chi$, so
we need to account for the phase space corrections included in
Eq.~(\ref{cross}) but not the simpler Eq.~(\ref{crossapprox}).

The flux of internal bremsstrahlung gamma rays is proportional to the
positron production rate, which is determined from the 0.511 MeV
intensity.  Positron annihilation occurs either via direct
annihilation into two 0.511 MeV gamma rays, or via the formation of a
bound state called positronium.  A singlet state (para-positronium),
which decays to two 0.511 MeV gamma rays, is formed 25\% of the time,
while the triplet state (ortho-positronium), which decays to three
continuum gamma rays, is formed 75\% of the time.  The ratio of direct
versus positronium annihilation can be measured by comparing the
0.511 MeV line intensity to the continuum intensity.  It is customary
to define the positronium fraction~\cite{Brown87} as
\begin{equation}
f = \frac{2}{1.5 + 2.25 (\Phi_{511}/\Phi_{\rm cont})}\,.
\end{equation}
This fraction depends on the physical conditions of the gas at the
Galactic center, and observations suggest that $f= 0.93 \pm 0.04$,
implying that most positrons annihilate via
positronium~\cite{Kinzer01}.  This means that the dark matter
annihilation rate and hence the internal bremsstrahlung rate are about
4 times larger than the positron annihilation rate observed via the
0.511 MeV line.

The spectrum of internal bremsstrahlung gamma rays per 0.511 MeV 
gamma ray is therefore
\begin{equation}
\frac{dN_{\rm Br}}{dE} = \left[ \frac{f}{4}+(1-f)\right]^{-1} 
\times \frac{1}{2} \times \frac{1}{\sigma_{\rm tot}}
\frac{d \sigma_{\rm Br}}{dE}\,, 
\end{equation}
where we have accounted for the fraction of positrons that annihilate
into 0.511 MeV gamma rays, and that each annihilation produces two
0.511 MeV gamma rays.


{\bf Three Key Assumptions.---}
We now identify three key assumptions of the model of Boehm {\it et
al.}~\cite{Boehm04a}; violations of these assumptions act in the sense
of {\it strengthening} the internal bremsstrahlung constraint.

First, it is assumed that the positron diffusion length is small
compared to the size of the Galactic center region (this may not be
the case; see Ref.~\cite{Bertone04b}).  Then the spatial distributions
of dark matter annihilation and the internal bremsstrahlung gamma rays
would be the same as the observed spatial distribution of $e^+ e^-$
annihilation.  If this assumption were violated, then the flux per
steradian of internal bremsstrahlung gamma rays would be larger (and
would come from a smaller angular region).

Second, it is assumed that the positrons are brought to rest by
ionization losses relatively quickly, so that the rates of dark matter
annihilation and positron annihilation are in equilibrium.  If this
assumption were violated, then the positron annihilation rate today
would reflect the dark matter annihilation rate in the past, while the
internal bremsstrahlung rate would reflect the (larger) dark matter
annihilation rate now.

Third, it is assumed that the positrons annihilate at rest.  If this
assumption were violated, then to produce the same 0.511 MeV flux, the
dark matter annihilation and internal bremsstrahlung gamma-ray rates
would have to be larger.  Additionally, positron annihilation in
flight could produce a detectable high-energy gamma-ray signature of
their own, though it would depend on the details of the physical
conditions in the Galactic center.


\begin{figure}[t]
\includegraphics[width=3.25in,clip=true]{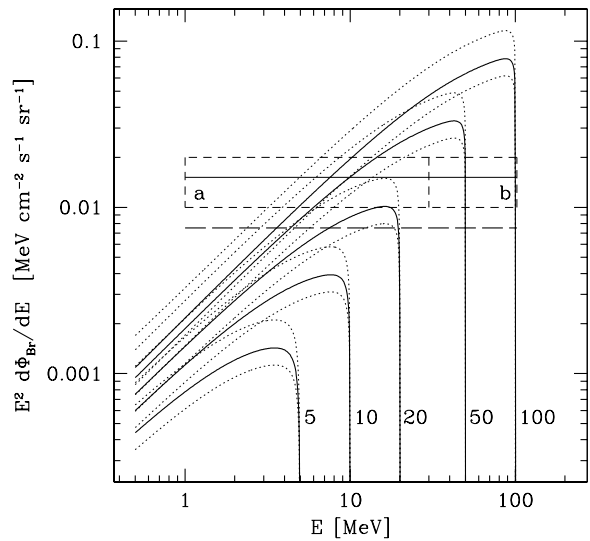}
\caption{\label{bremspectrum} The solid lines show the internal
bremsstrahlung spectra labeled by the assumed dark matter masses in
MeV, and normalized to produce the observed Galactic 0.511 MeV flux
(with the light dotted lines corresponding to its uncertainty); we
assumed that the FWHM of the 0.511 MeV emission region is 9 degrees.
The approximate COMPTEL (a) and EGRET (b) data and their uncertainty
band are shown by the box, and the long-dashed line below it indicates
the rough constraint on the maximum allowed contribution of internal
bremsstrahlung gamma-rays.}
\end{figure}

{\bf COMPTEL/EGRET Constraints.---}
The flux of internal bremsstrahlung gamma rays per steradian is
\begin{equation}
\frac{d\Phi_{\rm Br}}{dE} \simeq
\frac{1}{2} \Phi_{511} \frac{dN_{\rm Br}}{dE} \frac{1}{\Delta \Omega}\,,
\end{equation}
where we have assumed that $1/2$ of the 0.511 MeV flux is
emitted from an angular region $\Delta \Omega$ given by the gaussian
FWHM of 9 degrees.  This becomes
\begin{equation}
\frac{d\Phi_{\rm Br}}{dE}\simeq
\left(4.2^{+2.0}_{-0.9} \right) \times 10^{-2} 
\frac{1}{\sigma_{\rm tot}} \frac{d \sigma_{\rm Br}}{dE}
\, {\rm cm}^{-2} \, {\rm s}^{-1}\, {\rm sr}^{-1}\,.
\end{equation}
The corresponding internal bremsstrahlung gamma-ray spectra are shown
in Fig.~2.  The energy dependence of the cross section in
Eq.~(\ref{cross}) gives spectra which peak at $E\simeq m_\chi$, and
fall off as $1/E$; for easier comparison with the data, we plot $E^2$
times the spectra.  The offsets between the solid lines at low E
reflect just the $\ln \left( 2 m_\chi / m_e \right)$ factor in the
approximate and process-independent Eq.~(\ref{crossapprox}).  The mild
process dependence (the difference between Eq.~(\ref{cross}) and
Eq.~(\ref{crossapprox}) in the treatment of the phase space) is seen
in Fig.~2 in the slight turnovers in the solid lines before the
endpoints, as well as a general reduction by a factor $\lesssim 2$.
Had we used just Eq.~(\ref{crossapprox}), our constraints would have
been somewhat too strong.

We can constrain the internal bremsstrahlung contribution by comparing
to COMPTEL and EGRET measurements of the {\it diffuse} gamma-ray flux
from the Galactic center region~\cite{Strong98b,Strong98c,Strong04}.
Those fluxes were averaged over a Galactic center region of $\pm 5$
degrees in latitude and $\pm 30$ degrees in longitude, but were shown
to have mild variation across that region.  The observed 0.511 MeV
emission is from a smaller region of FWHM 9 degrees.  Thus in order to
constrain an internal bremsstrahlung contribution to the central
circle, we must first subtract the diffuse astrophysical contribution
as measured over the whole COMPTEL/EGRET rectangle.  We have
conservatively assumed that the COMPTEL/EGRET data band is given at 1
sigma, and that an excess contribution of more than about 50\% could
not be tolerated.  A fit to the full energy and angular dependence of
the signal and background would yield a more stringent constraint.

Even with our very conservative treatment, we can easily constrain the
dark matter mass in the Boehm {\it et al.} model~\cite{Boehm04a} to
$m_\chi \lesssim 20$ MeV (for decaying dark matter
~\cite{Picciotto04,Hooper04} we determine a similar upper limit).  The
uncertainties on the COMPTEL/EGRET data (as well as our approximate
handling thereof), the 0.511 MeV flux, and the positronium fraction
are relatively unimportant, though improvements would be welcomed.
The largest uncertainty on the constraint arises from the uncertainty
in the size of the 0.511 MeV emission region.  Following
Refs.~\cite{Jean03,Knodlseder03,Weidenspointner04,Boehm04a}, we
assumed 9 degrees, but the $2-\sigma$ range spans $6-18$ degrees,
allowing an order of magnitude in the internal bremsstrahlung flux per
steradian; this corresponds to varying the upper limit on $m_\chi$
from 10 to 60 MeV.  Improved results from INTEGRAL on the size of the
emission region are thus eagerly awaited.


{\bf Discussion and Conclusions.---}
In order to explain the Galactic positron excess, as well as the
smooth and centrally symmetric morphology of the 0.511 MeV emission
observed by INTEGRAL~\cite{Jean03,Knodlseder03,Weidenspointner04},
Boehm {\it et al.}~\cite{Boehm04a} (see also
Refs.~\cite{Boehm02,Boehm03,Hooper03,Boehm04b,Serpico04}) proposed an
intriguing model where $e^+ e^-$ pairs are produced by the
annihilation of light ($1-100$ MeV) dark matter candidates, which
apparently can evade all present accelerator and astrophysical
constraints.  We have pointed out the central assumption of this
model, i.e., that energetic positrons may be produced in the Galactic
center with no other observational consequences other than their
eventual annihilation into 0.511 MeV gamma rays.  We have shown that
such an ``invisible" birth is prohibited by the emission of internal
bremsstrahlung gamma rays from the original dark matter annihilations
unless the dark matter mass is less than about 20 MeV, disallowing
most of the proposed range.  We have arrived at this constraint in a
very conservative fashion, and expect that improved gamma-ray data and
a more sophisticated analysis will significantly improve the
sensitivity.  Our constraint is very nearly independent of the new
physics of the dark matter particles, and would be {\it strengthened}
by relaxing the assumptions on the physical conditions at the Galactic
center.

Although our results have been presented in the context of dark matter
annihilation, they are much more general.  {\it Any} mechanism which
produces energetic positrons will be accompanied by internal
bremsstrahlung gamma rays, and the rate is nearly independent of the
tree level cross section.  Thus for any mechanism that creates enough
positrons to account for the 0.511 MeV line, if those positrons are
produced above about 20 MeV, the accompanying internal bremsstrahlung
will violate the COMPTEL/EGRET constraints.  Dark matter decay is an
example of such a mechanism~\cite{Picciotto04,Hooper04}.  The ultimate
sensitivity of the technique at low energies will be reached when the
dominant sources are astrophysical sites which produce gamma rays and
positrons at comparable (tree-level) rates.


{\bf Acknowledgments.---} 
We are grateful to C.~Boehm, S.~Boggs, E.~Braaten, V.~Cirigliano,
D.~Hartmann, M.~Leising, D.~Rainwater and M.~Ramsey-Musolf for
comments on the manuscript.  JFB was supported by funds from The Ohio
State University, and NFB and GB by Fermilab (operated by URA under
DOE contract DE-AC02-76CH03000) and by NASA grant NAG5-10842.


\end{document}